\documentclass[aps,prb,twocolumn,groupedaddress,showpacs]{revtex4}
\usepackage{amssymb}
\usepackage[fleqn]{amsmath}
\usepackage[dvips]{graphicx}
\usepackage{docs}
\usepackage{bm}
\usepackage[colorlinks=true,linkcolor=blue,citecolor=blue,urlcolor=black]{hyperref}%
\expandafter\ifx\csname package@font\endcsname\relax\else
 \expandafter\expandafter
 \expandafter\usepackage
 \expandafter\expandafter
 \expandafter{\csname package@font\endcsname}%
\fi

\usepackage[papersize={9in,12in}]{geometry}
\newlength{\Figwidth}
\begin{document}
\title{The Electrical and Spin Properties of Monolayer and Bilayer Janus HfSSe under Vertical Electrical Field }

\author{Nayereh Ghobadi}
\affiliation{Department of Electrical Engineering, University of Zanjan, Zanjan, Iran}
\author{Shoeib Babaee Touski}\email{touski@hut.ac.ir}
\affiliation{Department of Electrical Engineering, Hamedan University of Technology, Hamedan, Iran}

\begin{abstract}
In this paper, the electrical and spin properties of mono- and bilayer HfSSe in the presence of a vertical electric field are studied. The density functional theory is used to investigate their properties. Fifteen different stacking orders of bilayer HfSSe are considered. The mono- and bilayer demonstrate an indirect bandgap, whereas the bandgap of bilayer can be effectively controlled by electric field. While the bandgap of bilayer closes at large electric fields and a semiconductor to metal transition occurs, the effect of a normal electric field on the bandgap of the monolayer HfSSe is quite weak. Spin-orbit coupling causes band splitting in the valence band and Rashba spin splitting in the conduction band of both mono- and bilayer structures. The band splitting in the valence band of the bilayer is smaller than a monolayer, however, the vertical electric field increases the band splitting in bilayer one. The stacking configurations without mirror symmetry exhibit  Rashba spin splitting which is enhanced with the electric field.

\end{abstract}



\maketitle

\section{introduction}
Two-dimensional (2D) transition-metal dichalcogenides (TMDs) attract intensive attention due to their extraordinary electronic, optical, mechanical and thermal properties \cite{manzeli20172d,choi2017recent}. In the layered TMDs with the chemical formula MX$_2$ (M =Mo, W, Zr, Hf; X = S, Se, Te), the transition metal atom M is sandwiched between two chalcogen atoms X to form X-M-X configuration by strong intra-layer chemical bonding. The layers are connected with weak van der Waals forces. 2D monolayer TMDs demonstrate rich physics and potential for broad applications in the fields of electronics, optoelectronics, catalysis, and spintronics \cite{manzeli20172d,touski2017enhanced}. TMD materials have been extensively studied due to their outstanding properties such as high carrier mobility and on/off current ratio. Most 2D TMD materials are semiconductors with a bandgap of about 2 eV which can be controlled by electric field \cite{kuc2015electronic, liu2017electric, sharma2014strain}, making them suitable for application in nanoelectronics and optoelectronics.

The bulk HfX$_2$ (Hafnium dichalcogenides) (X= S, Se) are TMD semiconductors with an indirect bandgap \cite{wu2017strain,reshak2005ab}. These compounds with quasi-layered structures stay together with low van der Waals force. The monolayer HfX$_2$ can be exfoliated by mechanical and chemical techniques \cite{wu2017strain,reshak2005ab}. Hafnium dichalcogenides have drawn considerable attention because of its fascinating electronic and optical properties
\cite{xu2016toward,tsipas2015epitaxial,salavati2019electronic}. Particularly, these materials demonstrate high mobility (1800 to 3500$cm^2/Vs$) and a large absorption range (1–2 eV). Monolayer HfS$_2$ and HfSe$_2$ are proposed as suitable materials for the electronic and optoelectronic applications  \cite{xu2015ultrasensitive, wang2018selective, yin2016ultrafast,kang2017tunable,yan2017space}. These materials demonstrate potential candidate for application in next-generation nanodevices such as water splitting and large scale solar cell \cite{singh20162d,wang2018recent}. Monolayer HfS$_2$ with a bandgap approximately equal to 2 eV, is theoretically confirmed to enhance the absorption portion of the sunlight for water splitting \cite{mattinen2019atomic}. The ultra-thin HfS$_2$ with high stability is demonstrated as phototransistors with high on/off ratio, high responsivity, small response time that shows remarkable potential for electronic and optoelectronic applications \cite{xu2015ultrasensitive}. Nowadays, numerous nanodevices based on the HfS$_2$ such as photodetectors and field-effect transistors have been experimentally fabricated \cite{zhang2019improvement,nie2017impact,wang2017epitaxial,wang2018selective}.

Bilayer ZrSe2 and HfSe2 demonstrate an indirect bandgap of 0.99 and 1.07 eV, respectively \cite{yan2019bilayer}. These two bilayers demonstrate high hole mobility which is one order of magnitude larger than the electron. The effect of the vertical electric field on the electronic properties of bilayer ZrS2 has been investigated. The bandgap of bilayer ZrS2 decreases with rising electric field \cite{shang2017effects} and semiconductor to metal transition occurs at a critical electrical field.

The electrical, and electronic properties of Junus ZrSSe monolayer are studied. It is confirmed that the ZrSSe monolayer has dynamically and mechanically stability. The ZrSSe monolayer is an indirect gap semiconductor and demonstrates amazing spin-orbit splitting \cite{guo2019predicted}. M. Barhoumi, et al \cite{barhoumi2020dft} studied the stability and electrical properties of monolayer and bilayer Janus HfSSe. The HfSSe bilayer is an indirect semiconductor, whereas, the HfSSe alternating in which the S and Se atoms are uniformly spread in the varying location, is a direct semiconductor. They studied four stacking orders of HfSSe bilayer, which display an indirect semiconductor with the bandgap in a range of 0.361-0.830 eV.

In this work, the electrical and spin properties of mono- and bilayer HfSSe are studied using density functional theory. Janus bilayer HfSSe with fifteen stacking orders is considered. Furthermore, the effect of the vertical electrical field on the electrical and spin properties is explored.

\begin{figure}[!t]
	\centering
	\includegraphics[width=1.0\linewidth]{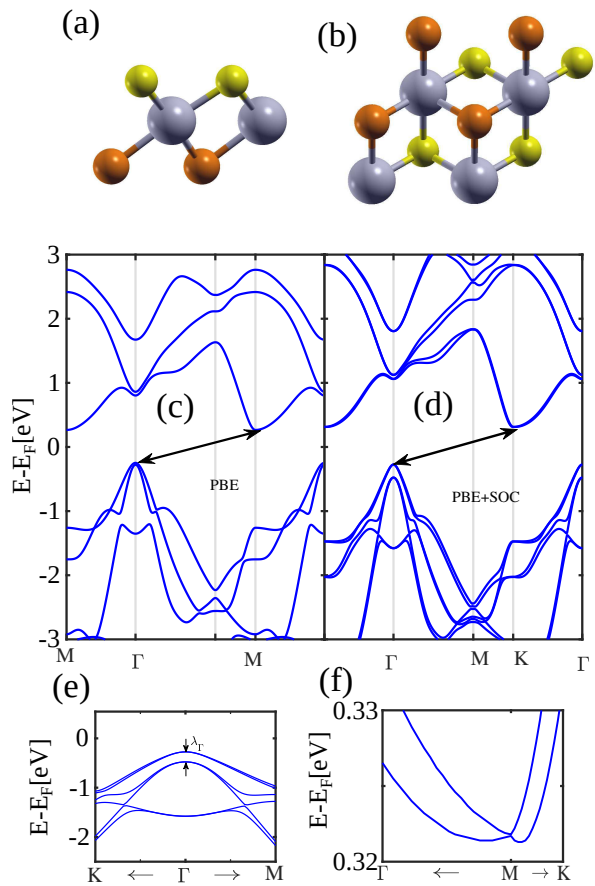}
	\caption{(a) Side and (b) Top view of the atomic structure of monolayer HfSSe. Band structure of monolayer HfSSe (c) without and (d) with SOC. The magnified view of the band structure around the (e) $\Gamma$-point in the valence band and (f) M-point in the conduction band.}
	\label{fig:fig1}
\end{figure}

\section{Computational details}
In order to investigate the properties of HfSSe, density functional calculations are performed using the SIESTA package \cite{soler2002siesta}. The generalized gradient approximation (GGA) with the Perdew-Burke-Ernzerhof (PBE) \cite{perdew1981self} functional is employed for the exchange-correlation term. We have adopted fully relativistic pseudopotentials and have taken into account the effect of Spin-orbit coupling (SOC). The van der Waals (vdW) interaction between layers in bilayer structures are treated using the Grimme's correction to the PBE functional \cite{grimme2006} which is the same method described in our previous study \cite{ghobadi2019normal,shamekhi2020band}. A Monkhorst-Pack k-point grid of $21\times21\times1$ is chosen for the unit-cell. The energy cutoff is set to be 50 Ry and a double-ζ$\zeta$ plus polarization basis-set is used. The total energy is converged to better than $10^{-5}$ eV and the geometries are fully relaxed until the force on each atom is less than 0.01 eV/$\AA$. A vacuum region of 30$\AA$  is added to avoid interactions in the normal direction. In order to visualize the atomic structures, XCrySDen package has been used \cite{kokalj2003}. A vertical electric field has been applied to the structures. The effective masses of the electrons and holes are calculated as \cite{touski2020electrical}:
\begin{equation}
m^*=\hbar^2/\left(\partial^2E/\partial k^2\right)
\end{equation}
where $\hbar$ is reduced Planck constant, $E$ and $k$ are the energy and wave vector of the conduction band minimum and the valence band maximum.

\begin{table}[t]
	\caption{The structural parameters (a and d$_{S-Se}$), bandgap (E$_g$), the band splitting at $\Gamma$ point ($\lambda_{\Gamma}$), Rashba coefficient ($\alpha_R$) and the effective masses of electron and hole in monolayer HfSSe. The length, effective mass, energy, and Rashba coefficient are in the units of $\AA$, $m_0$, eV, and $eV\AA$ respectively. \label{tab:tab1}}
	\begin{tabular}{p{1.3cm}p{1.2cm}p{1.2cm}p{1.5cm}p{1.1cm}}
		\hline
		\hline
		a & d$_{S-Se}$  & E$_g$ & $E_C-E_F$ & $\lambda_{\Gamma}$\\
		\hline
		3.718   & 3.043  & 0.587 & 0.313 &  0.203	\\
		\hline
		$m^{v,*}_{\Gamma,1}$ & $m^{v,*}_{\Gamma,2}$ & $m^{c,*}_{M\rightarrow \Gamma}$ & $m^{c,*}_{M\rightarrow K}$  & $\alpha_R$\\
		\hline
		0.212 & 0.191 & 2.116 & 0.197	& 0.159
		\\               
		\hline
		\hline
	\end{tabular}
\end{table}

\begin{figure*}
	\centering
	\includegraphics[width=1.0\linewidth]{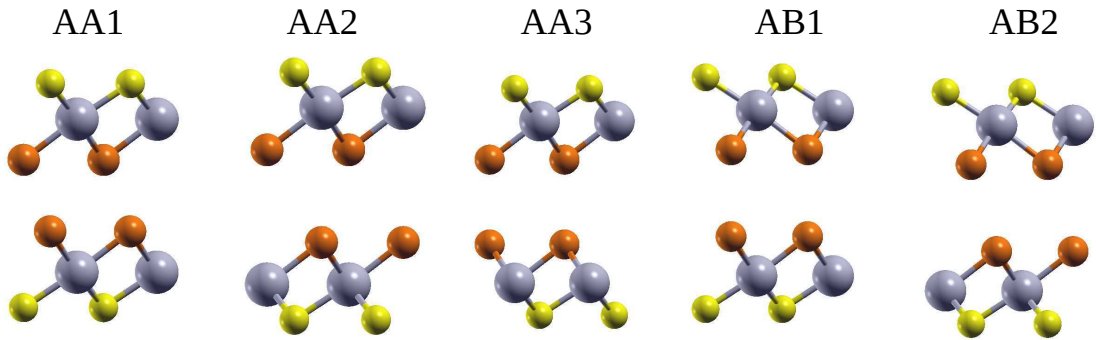}
	\caption{Side view of the atomic structures of different stacking orders of bilayer HfSSe. Fifteen stacking orders can be obtained by changing the position of S and Se atoms in a layer or both layers.}
	\label{fig:fig2}
\end{figure*}

\section{results and discussion}
The band structure of monolayer Janus HfSSe with and without spin-orbit coupling is depicted in Fig. \ref{fig:fig1}. The optimized structural, electrical, and spin properties of monolayer HfSSe are summarized in Table. \ref{tab:tab1}. The values of lattice constant and bandgap are calculated as 3.718$\AA$ and 0.587eV, receptively. The lattice constant is close to the previously reported values and the bandgap is almost the average value of the band gaps previously reported which are in the range of o.45 to 0.834 eV \cite{barhoumi2020dft,hoat2020comprehensive, shi2018mechanical, jena2018emergence}. Comparing Fig. \ref{fig:fig1} (c) and (d) one can conclude that SOC splits the degeneracy of first and second bands at $\Gamma$-point in the valence band. The value of this band splitting ($\lambda_{\Gamma}$) that represents the strength of spin-orbit coupling, is 0.203eV.  


HfSSe exhibits isotropic hole effective mass, whereas the electron effective mass is anisotropic. As one can see, the band structure is anisotropic around M-valley. The electron effective masses are 0.197m$_0$ and 2.116m$_0$ along M-K and M-$\Gamma$ paths, respectively. The first and second valence bands are close to each other at the valence band maximum (VBM). Therefore, the effective masses of first and second valence bands which are 0.212m$_0$ and 0.191m$_0$ should be considered. The electron effective masses are approximately close to HfS$_2$ and HfSe$_2$, whereas HfSSe demonstrates a lower hole effective mass \cite{yan2019bilayer,zhang2014two,iordanidou2016oxygen,iordanidou2016impact,kanazawa2016few,lu2018band}.

The minimum of the conduction band is placed at M-valley and the magnified view of the band structure around this point is depicted in Fig. \ref{fig:fig1}(f). As one can observe, the band structure displays Rashba splitting at M-valley. Rashba coefficient can be obtained as $\alpha_R=2E_R/K_R$ where $E_R$ and $K_R$ are Rashba energy and momentum, respectively. The value of the Rashba coefficient is negligible at M to $\Gamma$ path and we have reported it along with M to K path. The Rashba coefficient is 0.159$eV\AA$ that is smaller than reported values for other 2D materials  \cite{ariapour2020strain, absor2018strong, ma2014emergence, ariapour2019spin, li2017electronic, yao2017manipulation}. 
 
\begin{table}[t]
	\caption{The structural parameters of the fifteen stacking orders of bilayer HfSSe. The lattice constant (a), interlayer distance (d$_{int}$), the distance between Hf atoms of two layers (d$_{Hf-Hf}$), the distance between S and Se atoms in a layer (d$_{S-Se}$) are in $\AA$ unit and the binding energy ($E_b$) is in eV.	\label{tab:tab2}}
	
	\begin{tabular}{p{0.9cm}p{0.8cm}p{1.cm}p{0.9cm}p{1.2cm}p{1.cm}p{0.9cm}}
		\hline
		\hline
		Stacking order & &a & d$_{int}$  & d$_{Hf-Hf}$ & d$_{S-Se}$ & E$_b$ \\
		\hline
		AA1 & S-S & 3.684 & 2.8 & 5.607 & 3.046 & -2.958    \\
		& S-Se & 3.682 & 2.802 & 5.843 & 3.059 & -2.974    \\
		& Se-Se & 3.679 & 2.868 & 6.138 & 3.058 & -2.974    \\
		\hline
		AA2 & S-S & 3.688 & 2.961 & 5.767 & 3.047 & -2.889    \\
		& S-Se & 3.684 & 2.888 & 5.926 & 3.061 & -2.915    \\
		& Se-Se & 3.682 & 3.016 & 6.292 & 3.064 & -2.931    \\
		\hline
		AA3 & S-S & 3.685 & 3.672 & 6.483 & 3.052 & -2.698    \\
		& S-Se & 3.68 & 3.56 & 6.599 & 3.062 & -2.722    \\
		& Se-Se & 3.667 & 3.066 & 6.403 & 3.116 & -2.936    \\
		\hline
		AB1 & S-S & 3.685 & 3.682 & 6.498 & 3.05 & -2.697    \\
		& S-Se & 3.676 & 3.312 & 6.365 & 3.065 & -2.74    \\
		& Se-Se & 3.677 & 3.435 & 6.713 & 3.065 & -2.754    \\
		\hline
		AB2 & S-S & 3.687 & 2.809 & 5.614 & 3.049 & -2.937    \\
		& S-Se & 3.683 & 2.848 & 5.882 & 3.061 & -2.934    \\
		& Se-Se & 3.682 & 2.932 & 6.206 & 3.062 & -2.949    \\
		\hline
		\hline
	\end{tabular}
\end{table}

\begin{figure*}
	\centering
	\includegraphics[width=0.95\linewidth]{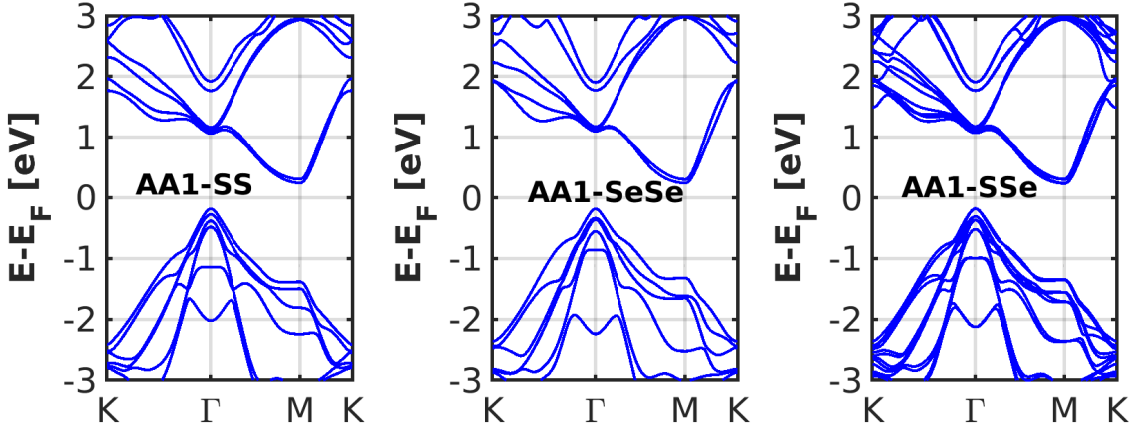}
	\caption{The band structure of AA1 stacking with SS, SeSe and SSe vertical configurations.}
	\label{fig:fig3}
\end{figure*}

\begin{table*}[t]
	\caption{The electrical and spin properties of the fifteen stacking orders of bilayer HfSSe. The bandgap ($E_g$), the energies of conduction ($E_c-E_F$) and valence band edges ($E_v-E_F$) and  the band splitting at $\Gamma$-point in the valence band ($\lambda_{\Gamma}$) are in eV unit. The effective masses at $\Gamma$-point for first and second valence band and at M-valley in the conduction band are in m$_0$ unit. The Rashba coefficient at M-valley in the conduction band ($\alpha_R$) is in $eV\AA$ unit.	 \label{tab:tab3}}
	\begin{tabular}{p{1.5cm}p{1.cm}p{1.2cm}p{1.7cm}p{1.7cm}p{1.3cm}p{1.3cm}p{1.3cm}p{1.2cm}p{1.2cm}p{1.3cm}}
		\hline
		Stacking order & & E$_g$ & E$_v$-E$_F$ & E$_c$-E$_F$ &$m^{v,*}_{\Gamma,1}$ & $m^{v,*}_{\Gamma,2}$ & $m^{c,*}_{M\rightarrow K}$ & $m^{c,*}_{M\rightarrow \Gamma}$  & $\lambda_{\Gamma}$ & $\alpha_R$ \\
		\hline
		\hline
		& S-S & 0.426 & -0.185 & 0.241 & 0.182 & 0.365 & 0.184 & 1.095 & 0.089 & 0\\
		AA1	& S-Se & 0.415 & -0.185 & 0.242 & 0.193 & 0.193 & 0.22 & 3.275 & 0.129 & 0.0925\\
		& Se-Se & 0.426 & -0.18 & 0.235 & 0.205 & 0.219 & 0.184 & 1.639 & 0.153 & 0\\
		\hline
		& S-S & 0.27 & -0.148 & 0.198 & 0.182 & 0.193 & 0.184 & 1.64 & 0.067 & 0\\
		AA2	& S-Se & 0.264 & -0.113 & 0.157 & 0.172 & 0.204 & 0.184 & 3.268 & 0.149 & 0.1211\\
		& Se-Se & 0.345 & -0.106 & 0.158 & 0.193 & 0.234 & 0.184 & 3.274 & 0.129 & 0\\
		\hline
		& S-S & 0.405 & -0.188 & 0.244 & 0.191 & 0.313 & 0.221 & 3.439 & 0.035 & 0\\
		AA3	& S-Se & 0.387 & -0.18 & 0.225 & 0.182 & 0.182 & 0.157 & 1.636 & 0.064 & 0.1429     \\
		& Se-Se & 0.347 & -0.169 & 0.219 & 0.193 & 0.193 & 0.184 & 1.642 & 0.13 & 0     \\
		\hline
		& S-S & 0.376 & -0.154 & 0.206 & 0.173 & 0.183 & 0.183 & 1.645 & 0.035 & 0     \\
		AB1	& S-Se & 0.292 & -0.166 & 0.21 & 0.183 & 0.194 & 0.183 & 1.645 & 0.123 & 0.0924     \\
		& Se-Se & 0.36 & -0.12 & 0.173 & 0.173 & 0.194 & 0.158 & 3.29 & 0.102 & 0     \\
		\hline
		& S-S & 0.405 & -0.16 & 0.212 & 0.182 & 0.252 & 0.184 & 1.093 & 0.092 & 0     \\
		AB2	& S-Se & 0.385 & -0.178 & 0.245 & 0.193 & 0.205 & 0.184 & 3.275 & 0.119 & 0.1457     \\
		& Se-Se & 0.378 & -0.171 & 0.226 & 0.205 & 0.205 & 0.158 & 1.638 & 0.168 & 0     \\
		\hline
		\hline
	\end{tabular}
\end{table*}

In the following, bilayer HfSSe is explored and its properties are discussed. The five stacking orders are depicted in Fig. \ref{fig:fig2}. In the vertical direction, there are three kinds of configurations for each stacking order: SeHfS/SHfSe, SeHfS/SeHfS, and SHfSe/SeHfS. We label these three vertical configurations as
SS, SSe, and SeSe. Therefore, considering the three vertical configurations for each stacking order of Fig. \ref{fig:fig2}, fifteen different stackings are obtained.

 The band structures of these fifteen configurations are similar with some differences. As a sample, we consider AA1 stacking which has the lowest energy and is the most stable structure. The band structures of AA1 stacking with SS, SeSe, and SSe vertical configurations are shown in Fig. \ref{fig:fig3}. The band structures are similar to monolayer HfSSe and exhibit an indirect bandgap. The valence band maximum and conduction band minimum are placed at $\Gamma$ and M-points, respectively. Up- and down-spin possess the same energies in SS and SeSe stacking, whereas, spin splitting occurs in SSe configuration. Rashba property is also observed in SSe stacking and other stackings don't display any Rashba splitting.
 

\begin{figure}
	\centering
	\includegraphics[width=1.0\linewidth]{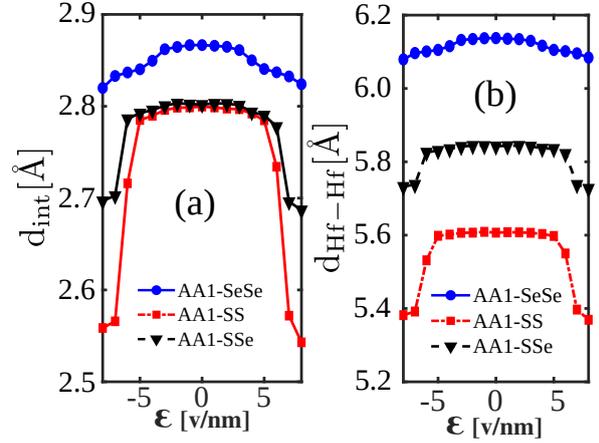}
	\caption{(a) Interlayer and (b) Hf-Hf distance of AA1 stacking with SS, SeSe and SSe vertical configurations as a function of electric field.}
	\label{fig:fig4}
\end{figure}

The structural properties of the fifteen stackings are listed in Table. \ref{tab:tab2}. The lattice constants are approximately the same for all stackings and are close to the lattice constant of the monolayer. The interlayer distance ($d_{int}$) is in the range of 2.8 to 3.682 $\AA$. The distance between Hf atoms of two layers ($d_{Hf-Hf}$) and the vertical distance between S and Se atoms in a layer ($d_{S-Se}$) are also reported. AA1 stacking has the lowest binding energy regardless of the vertical configuration, indicating that the  AA1 structure is the most stable stacking order.


The electrical and spin properties of the bilayer HfSSe are summarized in Table. \ref{tab:tab3}. The bandgap is in the range of 0.264 to 0.426 eV that is lower than a monolayer. AA1 and AA2 stackings possess the highest and lowest bandgap, respectively. Both conduction and valence band edges are also closer to the Fermi level compared to monolayer ones. The effective masses for conduction and valence bands are listed in Table. \ref{tab:tab3}. As one can observe, the effective masses are approximately close to monolayer where some stackings display higher and others lower values. The valence band splitting at $\Gamma$-point is in the range of 35 to 160 meV. Finally, the SSe vertical configurations display Rashba splitting due to the lack of mirror symmetry, whereas, two SS and SeSe configurations demonstrate mirror symmetry. In the following, we consider AA1 configurations due to their lower energies, whereas, similar behavior has been observed in the other stackings.

The vertical electric field is known as a useful method to tune electrical properties. The effect of vertical electric field on the electrical and spin properties of mono- and bilayer HfSSe is explored. Interlayer distance and distance between Hf atoms of two layers as a function of the electric field are plotted in Fig. \ref{fig:fig4}. $d_{Hf-Hf}$ and $d_{int}$ in SS and SSe configurations are approximately unchanged at small values of the electric field, whereas they display a decreasing behavior under larger values of the electric field. On the other hand, $d_{Hf-Hf}$ and $d_{int}$ in SeSe stacking demonstrates an almost monotonous decreasing pattern under the electric field.

\begin{figure}
	\centering
	\includegraphics[width=1.0\linewidth]{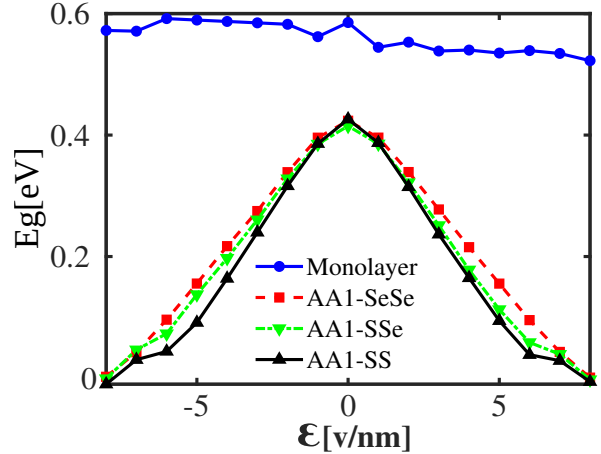}
	\caption{The band gap for mono- and bilayer HfSSe as a function of electric field. AA1 stackings are considered. }
	\label{fig:fig5}
\end{figure}

\begin{figure}
	\centering
	\includegraphics[width=1.0\linewidth]{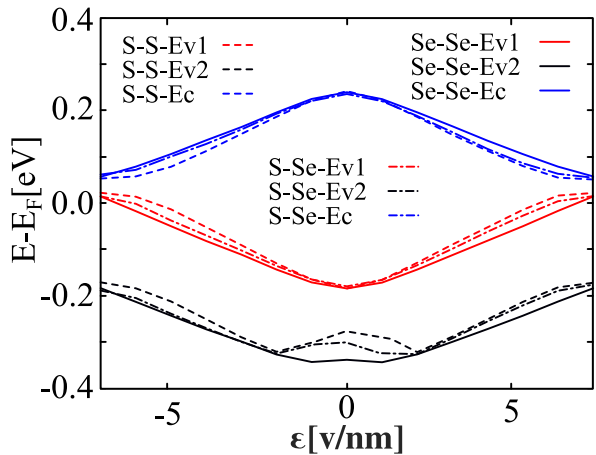}
	\caption{The variation of conduction, first and second valence bands in AA1 stackings as a function of electric field. }
	\label{fig:fig6}
\end{figure}

The bandgap versus electric field in mono- and bilayer are shown in Fig. \ref{fig:fig5}. As one can see, the effect of the electric field on the bandgap of the monolayer structure is quite weak. It slowly decreases with an increase in the electric field. On the other hand, the bandgap of bilayer decreases with the negative and positive electric field. Such behavior also reported in the bilayer ZrS$_2$ \cite{shang2017effects}. In addition, it has been reported that the bandgap of monolayer ZrSSe shows a little dependency on the vertical electric field \cite{vu2019electronic}. At equilibrium condition, AA1 configurations have approximately the same bandgap values, see Table. \ref{tab:tab2}. With applying the electric field, AA1-SS stacking shows a steeper decreasing pattern compared to SeSe and SSe structures. Eventually, the bandgap of three configurations close at a field strength of $\pm$7V/nm. So, we have plotted all figure in this range.

The effective masses and the energies of conduction and valence band edges are also investigated with the variation of the electric field. The effective masses don't show any changes and remain constant under the vertical electric field. On the other hand, the value of the conduction band minimum (CBM) decreases and the valence bands maximum (VBM) increases with the electric field, see Fig. \ref{fig:fig6}. The second valence band displays a local maximum at a field strength of 0 V/nm. This band decreases under small electric fields and then increases for larger fields. The second valence band maximum in S-S stacking demonstrates the highest local maximum and Se-Se owns the smallest. These results indicate that S-S stacking possesses the highest bonding and the interlayer bonding decreases in Se-Se stacking.

\begin{figure}
	\centering
	\includegraphics[width=1.0\linewidth]{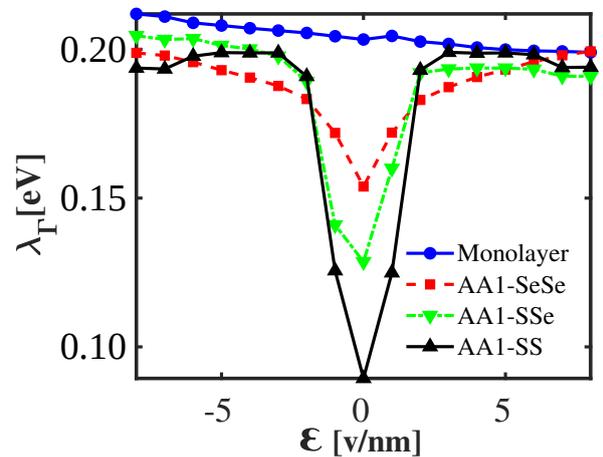}
	\caption{Valence band splitting at $\Gamma$-point for mono and bilayer HfSSe for various electric field. }
	\label{fig:fig7}
\end{figure}


The energy difference between the first and second valence band edges determines the contribution of the second band on the hole properties. The Band splitting at $\Gamma$-point in the valence band of mono- and bilayer under various electric fields are compared in Fig. \ref{fig:fig7}. Band splitting of the monolayer is larger than that of the bilayer and decreases smoothly with the electric field. Bilayer configurations demonstrate small $\lambda_{\Gamma}$ at equilibrium condition that increases sharply at low electric fields and changes slowly at larger electric fields. In the SS and SSe structures with mirror symmetry, the variation of $\lambda_{\Gamma}$ at large negative and positive electric fields is symmetric, whereas, in SSe structure the $\lambda_{\Gamma}$ increases smoothly at negative fields and decrease slowly at positive fields due to the lack of mirror symmetry.

\begin{figure}
	\centering
	\includegraphics[width=1.0\linewidth]{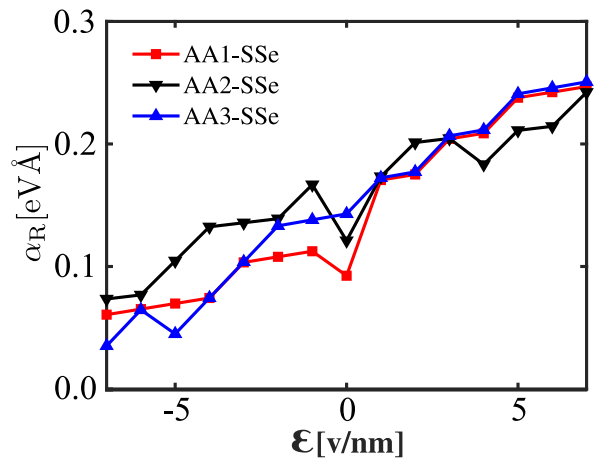}
	\caption{The Rashba coefficient versus electric field for SSe vertical configuration of AA stackings. }
	\label{fig:fig8}
\end{figure}

Next, we have investigated the influence of the vertical electrical field on the Rashba coefficient in polar HfSSe and the results are plotted in Fig. \ref{fig:fig8}. It has been reported in Table \ref{tab:tab2} that only SSe stackings demonstrate Rashba splitting. We applied the electrical field to all stackings and we found that all stackings demonstrate Rashba splitting under vertical electric field, but SSe stackings display a considerable Rashba coefficient so the Rashba coefficient is only plotted for SSe stacking. As one can observe, the Rashba coefficient variation of all stackings are close to each other, therefore, $\alpha_R$ is not affected by stacking order. The Rashba parameter increases with increasing electric field for all stackings. Such an increasing behavior of $\alpha_R$ with electric field is previously reported in other 2D materials \cite{ju2020electric,hu2018intrinsic,cheng2016nonlinear}. In the SSe stackings there exist an internal electric field and dipole moment which is due to the lack of mirror symmetry. When a positive electric field is applied, the internal electric field is strengthened, whereas that is weakened when a negative electric field is applied.

\section{Conclusion}
The electrical and spin properties of mono- and bilayer HfSSe in the presence of the vertical electrical field are studied. In monolayer structure, the value of band splitting at $\Gamma$-point in the valence band is 0.203eV, and the Rashba coefficient at M-valley in the conduction band is 0.159$eV\AA$. Fifteen different stacking orders of bilayer HfSSe are studied and a vertical electric field is applied to all. The electric field decreases interlayer distance and the bandgap closes at $\pm$7V/nm in the bilayer, whereas, has a negligible effect on the bandgap of the monolayer. Although the band splitting of bilayer configurations at $\Gamma$-point is smaller than that of the monolayer, the electric field increases it in bilayer structures so that it reaches to band splitting of monolayer under a large electric field. SSe vertical configurations display Rashba splitting in equilibrium condition which is comparable to the monolayer. However the electrical field generates Rashba splitting in all stacking orders, but the Rashba coefficient only is considerable for SSe configurations and is enhanced with the increasing vertical electric field.



\end{document}